\documentclass[aps,prc,amsmath,preprint,superscriptaddress,showpacs]{revtex4}
\usepackage{bm,graphicx}
\begin{document}

%\preprint{}

\title{Cross-sections and analyzing powers for 
$(p,n)$ reactions on ${}^{3}{\rm He}$ and 
${}^{4}{\rm He}$ at 346 MeV}


%\thanks{}
\author{E.~Ihara}
\affiliation{Department of Physics, Kyushu University,
Higashi, Fukuoka 812-8581, Japan}
\author{T.~Wakasa}
\email[]{wakasa@phys.kyushu-u.ac.jp}
\homepage[]{http://www.kutl.kyushu-u.ac.jp/~wakasa}
\affiliation{Department of Physics, Kyushu University,
Higashi, Fukuoka 812-8581, Japan}
\author{M.~Dozono}
\affiliation{Department of Physics, Kyushu University,
Higashi, Fukuoka 812-8581, Japan}
\author{K.~Hatanaka}
\affiliation{Research Center for Nuclear Physics, Osaka University,
Ibaraki, Osaka 567-0047, Japan}
\author{T.~Imamura}
\affiliation{Department of Physics, Kyushu University,
Higashi, Fukuoka 812-8581, Japan}
\author{M.~Kato}
\affiliation{Research Center for Nuclear Physics, Osaka University,
Ibaraki, Osaka 567-0047, Japan}
\author{S.~Kuroita}
\affiliation{Department of Physics, Kyushu University,
Higashi, Fukuoka 812-8581, Japan}
\author{H.~Matsubara}
\affiliation{Research Center for Nuclear Physics, Osaka University,
Ibaraki, Osaka 567-0047, Japan}
\author{T.~Noro}
\affiliation{Department of Physics, Kyushu University,
Higashi, Fukuoka 812-8581, Japan}
\author{H.~Okamura}
\affiliation{Research Center for Nuclear Physics, Osaka University,
Ibaraki, Osaka 567-0047, Japan}
\author{K.~Sagara}
\affiliation{Department of Physics, Kyushu University,
Higashi, Fukuoka 812-8581, Japan}
\author{Y.~Sakemi}
\affiliation{Cyclotron and Radioisotope Center, Tohoku University,
Sendai, Miyagi 980-8578, Japan}
\author{K.~Sekiguchi}
\affiliation{RIKEN Nishina Center,
Wako, Saitama 351-0198, Japan}
\author{K.~Suda}
\affiliation{RIKEN Nishina Center,
Wako, Saitama 351-0198, Japan}
\author{T.~Sueta}
\affiliation{Department of Physics, Kyushu University,
Higashi, Fukuoka 812-8581, Japan}
\author{Y.~Tameshige}
\affiliation{Research Center for Nuclear Physics, Osaka University,
Ibaraki, Osaka 567-0047, Japan}
\author{A.~Tamii}
\affiliation{Research Center for Nuclear Physics, Osaka University,
Ibaraki, Osaka 567-0047, Japan}
\author{H.~Tanabe}
\affiliation{Department of Physics, Kyushu University,
Higashi, Fukuoka 812-8581, Japan}
\author{Y.~Yamada}
\affiliation{Department of Physics, Kyushu University,
Higashi, Fukuoka 812-8581, Japan}

\date{\today}

\begin{abstract}
 The cross-sections and analyzing powers for
$(p,n)$ reactions on ${}^{3}{\rm He}$ and ${}^{4}{\rm He}$
have been measured 
at a bombarding energy of $T_p$ = 346 MeV and 
reaction angles of $\theta_{\rm lab}$ =  $9.4^{\circ}$--$27^{\circ}$.
 The energy transfer spectra for ${}^{3}{\rm He}(p,n)$ at large 
$\theta_{\rm lab}$ ($\ge$ $16^{\circ}$)
are dominated by quasielastic contributions, and 
can be reasonably reproduced by plane-wave impulse 
approximation (PWIA) calculations for quasielastic scattering.
 By contrast, the known $L$ = 1 resonances in ${}^{4}{\rm Li}$
are clearly observed near the threshold 
in the ${}^{4}{\rm He}(p,n)$ spectra.
 Because these contributions are remarkable at small angles,
the energy spectra are significantly different from those
expected for quasielastic scattering.
The data are compared with the PWIA calculations, and 
it is found that the quasielastic contributions are dominant 
at large $\theta_{\rm lab}$ ($\ge$ $22^{\circ}$).
 The nuclear correlation effects on the quasielastic peak 
for ${}^{4}{\rm He}(p,n)$ are also discussed.
\end{abstract}

%\keywords{}

\maketitle

\section{INTRODUCTION}
\label{sec:intro}

 In this article, we present the cross-sections and 
analyzing powers for $(p,n)$ reactions 
on ${}^{3}{\rm He}$ and ${}^{4}{\rm He}$ at a proton 
incident energy of $T_p$ = 346 MeV and laboratory reaction
angles of $\theta_{\rm lab}$ = $9.4^{\circ}$--$27^{\circ}$.
 The quasielastic scattering data obtained at 
$T_p$ = 346 MeV is of particular 
interest since the experimental data 
including the present results 
can cover a wide range of 
nuclei from ${}^{2}{\rm H}$ to ${}^{208}{\rm Pb}$ at the 
same incident energy 
\cite{prc_59_3177_1999,prc_69_044602_2004,prc_69_054609_2004}.
 It should be noted that the distortion in the nuclear 
mean field is minimal for a nucleon kinetic energy of 
about 300 MeV.
 Thus, the experimental data 
will provide important information to test 
theoretical calculations for the quasielastic process.

 One of the unique features of the $(p,n)$ quasielastic 
reaction is that the observed peak of the quasielastic 
distribution is shifted at higher excitation energy of 
about 20 MeV than expected from free nucleon-nucleon
({\it NN}\,) kinematics \cite{prc_52_228_1995}.
 Discrepancies between experimental data and theoretical 
predictions based on the free {\it NN} interaction 
may arise, for example, 
from nuclear many-body effects \cite{prc_63_044609_2001}
or multi-step effects \cite{prc_65_064616_2002}.
 For the ${}^{2}{\rm H}(p,n)$ reaction, the peak position 
is consistent with the corresponding free {\it NN} value, and 
the quasielastic distribution is reasonably reproduced by
plane-wave impulse approximation (PWIA) calculations 
\cite{prc_69_044602_2004}.
 The multi-step effects for $A$ $\le$ 4 nuclei 
are expected to be much smaller than those 
for medium and heavy nuclei.
 Thus the data for $A$ $\le$ 4 systems give clear
information on the nuclear many-body effects including
nuclear correlations.
 Hence it is very interesting to investigate whether 
the ${}^{3}{\rm He}(p,n)$ data can be reproduced with 
first-order model calculations.
% The experimental data is compared with theoretical 
%calculations using the Fermi-gas response function 
%in order to assess the Pauli principle effects properly.

 In contrast to the ${}^{3}{\rm He}(p,n)$ reaction, 
the spectra of the ${}^{4}{\rm He}(p,n)$ reaction 
exhibit prominent resonances with angular 
momentum transfer $L$ = 1 near the threshold 
\cite{npa_541_1_1992}.
 In the measurements at $T_p$ = 100 and 200 MeV 
\cite{plb_368_39_1996,prc_58_645_1998,prc_69_024616_2004}, 
there is no distinct quasielastic peak in the 
measured spectra because the resonance contributions 
are dominant.
 These data have been compared with calculations obtained
using the quasielastic scattering code 
{\sc threedee} \cite{prc_58_645_1998} 
and with recoil-corrected continuum shell-model
(RCCSM) calculations \cite{plb_368_39_1996,prc_69_024616_2004}.
 Both calculations reproduce the $L$ = 1 resonance 
contributions qualitatively, however, quantitative 
reproduction could not be achieved for the 
$L=1$ resonance or quasielastic contributions.

 The present data are compared with the calculations 
for quasielastic scattering 
\cite{prc_63_044609_2001},
which have been used extensively to analyze
quasielastic scattering data measured at 
LAMPF 
\cite{prl_69_582_1992,prc_47_2159_1993,prl_73_3516_1994} 
and 
RCNP \cite{prc_59_3177_1999,prc_69_054609_2004}.
 The ${}^{3}{\rm He}(p,n)$ data at large angles
are reasonably reproduced by the PWIA calculations,
which certify the predominance of the quasielastic process
around the peak region and the weakness of the nuclear
correlations.
 The calculations slightly underestimate the cross-sections
at small angles, which might be due to the proposed
isospin $T$ = $3/2$ three-nucleon resonance 
\cite{prl_23_1181_1969,prc_77_054611_2008}.
 For ${}^{4}{\rm He}(p,n)$, the underestimation is significant at small angles because 
the $L$ = 1 resonances in ${}^{4}{\rm Li}$ could not 
be described in the present quasielastic formalism.
 At large angles, where the quasielastic process is 
dominant, a significant difference is observed between the 
experimental and theoretical results
for the peak positions.
 The large $Q$-value and nuclear correlation effects 
are investigated 
since these effects are expected to be important 
in ${}^{4}{\rm He}$ \cite{prc_49_789_1994}.
 It is found that the discrepancy could be resolved 
in part by the nuclear correlations.

\section{EXPERIMENTAL METHODS}
\label{sec:exp}

 The experiment was carried out using the 
West--South Beam Line (WS-BL) \cite{nim_a482_79_2002} 
at the Research Center for Nuclear Physics (RCNP), Osaka University.
 The beam line configuration and the doubly achromatic
beam properties have been reported previously \cite{nim_a482_79_2002}.
 In the following, therefore, we discuss experimental details 
relevant to the present experiment.

\subsection{Polarized proton beam}

 The polarized proton beam was produced by 
the high-intensity polarized ion source (HIPIS) at RCNP
\cite{nima_384_575_1997}.
 In order to minimize geometrical false asymmetries,
the nuclear polarization state was cycled 
every 5 s between the normal and reverse states 
(e.g., between the ``up'' and 
``down'' states at the target position) 
by selecting rf transitions.
 The beam was accelerated to $T_p$ = 346 MeV 
by using the AVF and ring cyclotrons.
 One out of seven beam pulses was selected before injecting into 
the Ring cyclotron.
 This pulse selection yielded a beam pulse period of 
431 ns, and reduced the wraparound of slow 
neutrons from preceding beam pulses.
 A single-turn extraction was maintained during the measurement 
in order to keep the beam pulse period.
 Multi-turn extracted protons were less than 1\% of single-turn 
extracted protons.

 The beam polarization was continuously monitored
with the beam-line polarimeter BLP1 \cite{nim_a482_79_2002}.
 The polarimeter consisted of four pairs of conjugate-angle 
plastic scintillators.
 The $\vec{p}+p$ elastic scattering was used as the analyzing 
reaction, and the elastically scattered and recoiled protons 
were detected in kinematical coincidence with a pair of scintillators.
 A self-supporting ${\rm CH_2}$ target with a 
thickness of 1.1 ${\rm mg/cm^2}$ was used as the hydrogen target.
 The typical magnitude of the beam polarization was about 0.52.

\subsection{${}^{3}{\rm He}$ and ${}^{4}{\rm He}$ targets}

 The ${}^{3}{\rm He}$ and ${}^{4}{\rm He}$ targets were 
prepared as high-pressure cooled gas targets 
\cite{prc_77_054611_2008}
by using a target system developed for a 
liquid ${\rm H}_2$ target \cite{mpla_18_322_2003}.
 This target was operated at temperatures down to 29 K and 
at absolute pressures up to 2.5 atm.
 Both the cell temperature and pressure were continuously 
monitored during the experiment, and the typical target 
densities were about 
$1.2\times 10^{21}$ ${\rm cm^{-2}}$ and 
$1.0\times 10^{21}$ ${\rm cm^{-2}}$ for ${}^{3}{\rm He}$ 
and ${}^{4}{\rm He}$, respectively.
 The gas cell windows were made of 12-$\mu {\rm m}$-thick 
Alamid foil.
 Background spectra were also measured 
by filling the target cell with ${\rm H}_2$ gas
in order to subtract the contributions from both the Alamid
windows and the beam ducts.
 We also measured data with ${\rm D}_2$ gas in the target 
cell to determine the detection efficiency of the 
neutron detector system.
 These data were also used to estimate the systematic 
uncertainty as described below because the cross-sections 
are reliably predicted by the theoretical calculations.

\subsection{Dipole magnet and neutron detector}

 A dipole magnet made of permanent NEOMAX magnets \cite{neomax} 
was installed 10 cm downstream from the target.
 This magnet had a magnetic rigidity of $B\rho$ = 0.95 Tm, which 
was sufficient to sweep charged particles from the target 
in order to prevent them from entering the neutron 
detector system.
 
\begin{figure}
\begin{center}
\includegraphics[width=0.7\linewidth,clip]{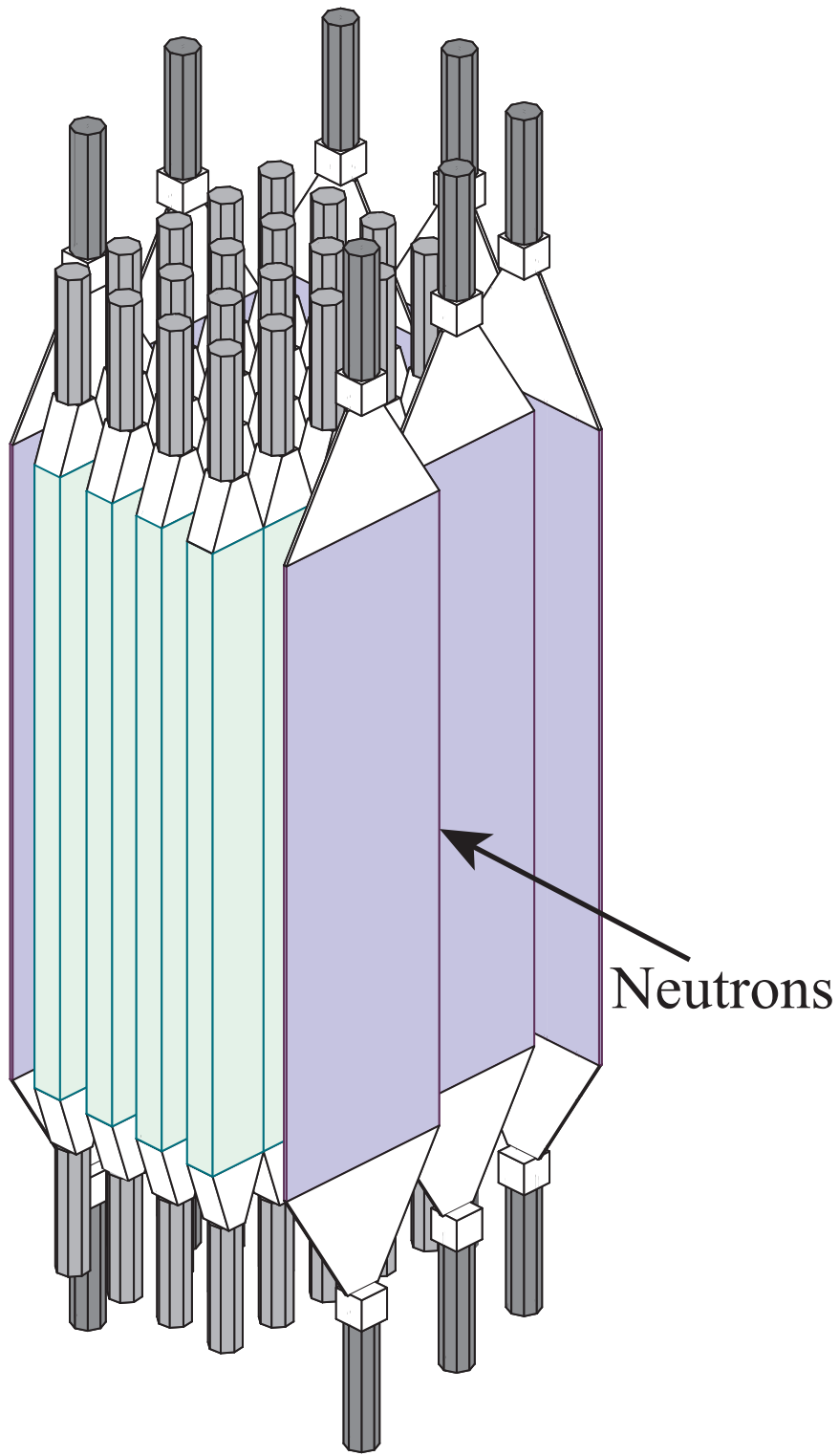}
\end{center}
\caption{(Color online)
 A schematic view of the neutron detector system.
 The 20 sets of neutron detectors are surrounded by
thin plastic scintillation detectors in order to reject 
charged particles.
}
\label{fig:detector}
\end{figure}

 Neutrons were measured with a 20~m flight path length
at $\theta_{\rm lab}$ = $9.4^{\circ}$--$27^{\circ}$.
 As illustrated in Fig.~\ref{fig:detector}, 
the neutron detector system consisted of 20 sets of 
one-dimensional position-sensitive plastic scintillators (BC408) 
with a size of 100 $\times$ 10 $\times$ 5 ${\rm cm}^3$, which was
part of the NPOL3 system \cite{nima_547_569_2005}.
 The detector system consisted of four planes of neutron detectors,
each with an effective solid angle of 
$\Delta \Omega$ = 1.25 msr.

\section{DATA REDUCTION}
\label{sec:reduction}

\subsection{Background subtraction}
 
\begin{figure}
\begin{center}
\includegraphics[width=0.7\linewidth,clip]{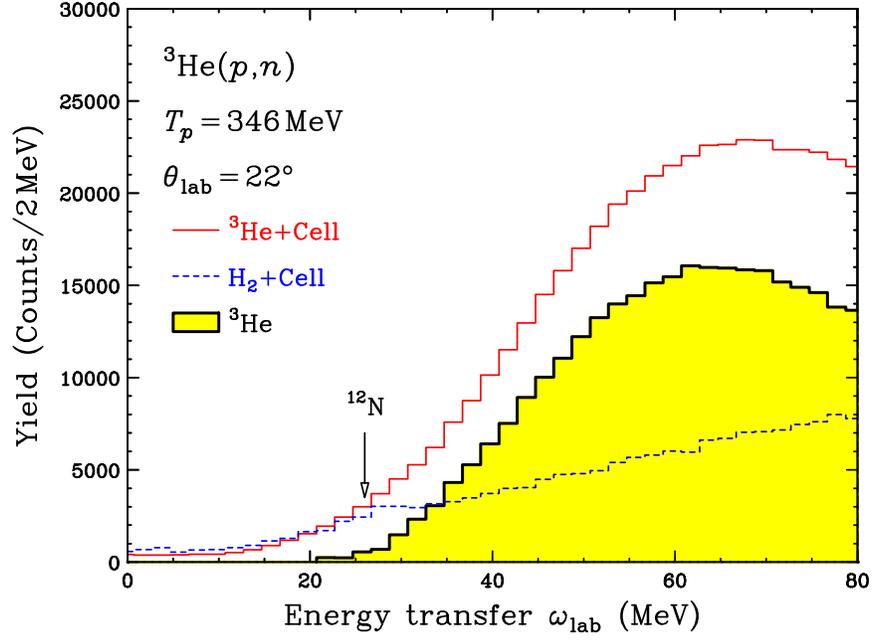}
\end{center}
\caption{(Color online)
 Energy transfer spectra for the targets 
filled with ${}^{3}{\rm He}$ (thin solid histogram) 
and ${\rm H}_2$ (dash histogram) gases 
for the $(p,n)$ reaction 
at $T_p$ = 346 MeV and $\theta_{\rm lab}$ = $22^{\circ}$.
 The filled thick solid histogram shows the spectrum
for the ${}^{3}{\rm He}(p,n)$ reaction obtained by the 
subtraction.}
\label{fig:sub}
\end{figure}

 The neutrons from the target windows
are the dominant source of the 
background at lower energy transfers 
of $\omega_{\rm lab}$ $\lesssim$ 50 MeV.
 This contribution can be subtracted by measuring 
the empty-target spectra. 
 However, the background from the beam ducts 
downstream from the target becomes significant at higher 
energy transfers.
 The proton beam was spread out by the multiple scattering 
in the target material, and part of it hit the beam ducts.
 This contribution depends on the target density, and thus 
we could not subtract it with the empty-target spectra. 
 Thus, we also measured the data with 
${\rm H}_2$ gas in the target cell because the multiple
scattering effects for He and ${\rm H}_2$ are expected to be similar.
 Figure~\ref{fig:sub} shows a representative set of 
spectra as a function of energy transfer.
 In the ${\rm H}_2$+Cell spectrum, 
a shoulder component is observed at 
$\omega_{\rm lab}$ $\simeq$ 26 MeV.
 This bump is mainly due to the spin-dipole resonances (SDRs)
in ${}^{12}{\rm N}$ excited by the $(p,n)$ reaction on 
${}^{12}{\rm C}$ in Alamid.
 The yield at $\omega_{\rm lab}$ $\gtrsim$ 30 MeV consists 
of the quasielastic $(p,n)$ reaction events 
for the Alamid windows and the background neutrons from the 
beam ducts.

 The filled histogram in Fig.~\ref{fig:sub} shows the 
subtraction results. 
 The background contributions including the SDR bump in 
${}^{12}{\rm N}$ are successfully subtracted 
without adjusting the relative normalization.
 We have also measured the data for the 
${}^{2}{\rm H}(p,n)$ reaction in order to 
investigate the reliability of the background subtraction.
 The results are discussed in the next section.

\subsection{Background subtracted observables}

 Observables for the $A(p,n)$ reaction ($A$ represents 
${}^{2}{\rm H}$, ${}^{3}{\rm He}$, or ${}^{4}{\rm He}$)
were 
extracted through a cross-section-weighted subtraction 
of the observables for the ${\rm H}_2$ target 
(${\rm H_2}+{\rm Cell}$) 
from the observables for the $A$ target 
($A+{\rm Cell}$)
as 
\begin{subequations}
\label{eq_sub}
\begin{eqnarray}
\sigma_{A} & = & 
   \sigma_{A+{\rm Cell}}-\sigma_{{\rm H_2}+{\rm Cell}}\ ,\\
D_{A} & = & 
\frac{D_{A+{\rm Cell}}-fD_{{\rm H_2}+{\rm Cell}}}{1-f}\ ,
\end{eqnarray}
\end{subequations}
where $\sigma$ represents the cross-section, 
$D$ is the analyzing power $A_y$,
and $f=\sigma_{{\rm H_2}+{\rm Cell}}/\sigma_{A+{\rm Cell}}$.
 The fraction $f$ was estimated by using the cross-sections
based on the nominal target thicknesses and integrated 
beam current.

\subsection{Neutron detection efficiency and energy resolution}

 The neutron detection efficiency was determined 
using the ${}^{2}{\rm H}(p,n)$ reaction at 
$\theta_{\rm lab}$ = $22^{\circ}$ 
whose cross-section at $T_p$ = 345 MeV is known 
\cite{prc_69_044602_2004}.
 The result is 0.035$\pm$0.002, 
where the uncertainty comes mainly from the 
uncertainty in 
the thickness of the ${}^{2}{\rm H}$ target.

 The overall energy resolution was determined 
by measuring the $p+p$ elastic scattering for the 
${\rm H}_2$ target.
 The result is $\Delta E$ = 7.7 MeV in full width 
at half maximum.

\section{RESULTS AND DISCUSSIONS}
\label{sec:results}

\subsection{${}^{2}{\rm H}(p,n)$ data}
\label{sec:res:2h}

\begin{figure}
\begin{center}
\includegraphics[width=0.7\linewidth,clip]{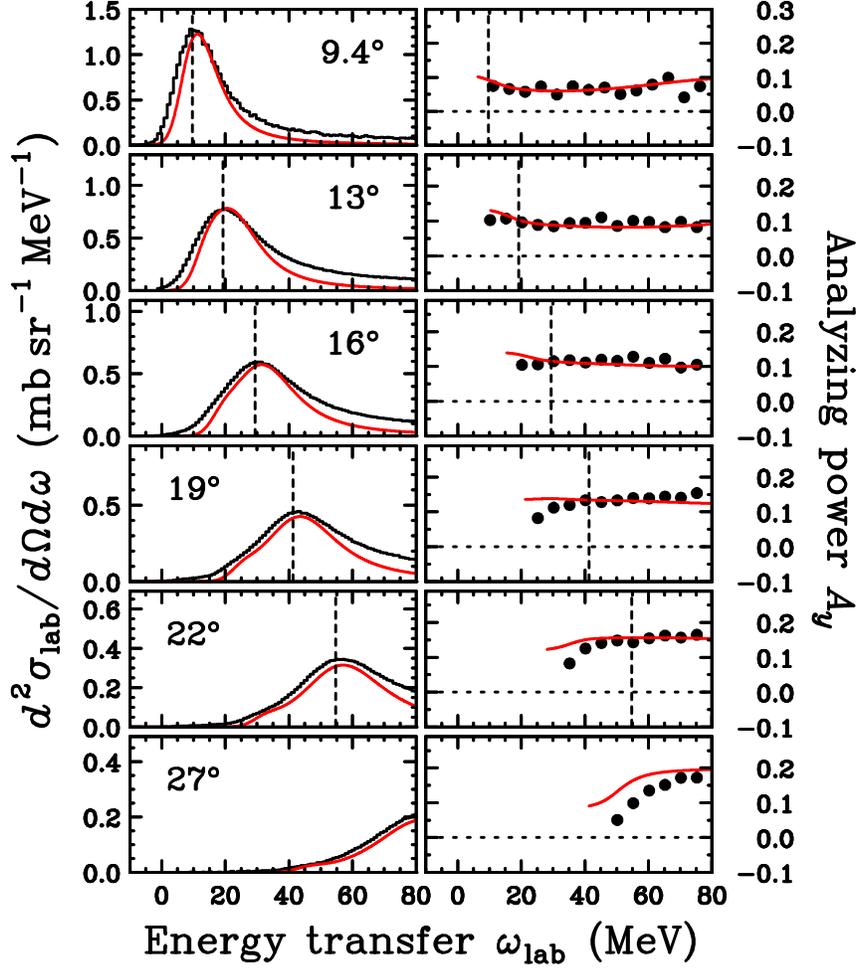}
\end{center}
\caption{(Color online)
 The cross-sections (left panels) and analyzing powers 
(right panels)
for the ${}^{2}{\rm H}(p,n)$ reaction
at $T_p$ = 346 MeV and $\theta_{\rm lab}$ = $9.4^{\circ}$--$27^{\circ}$.
 The vertical dashed lines represent the energy transfers 
for the free {\it NN} scattering.
 The solid curves are the PWIA predictions obtained by
the optimal factorization approximation.}
\label{fig:2h}
\end{figure}

 The cross-sections and analyzing powers for the 
${}^{2}{\rm H}(p,n)$ reaction at $T_p$ = 346 MeV and 
$\theta_{\rm lab}$=
$9.4^{\circ}$--$27^{\circ}$ are presented in Fig.~\ref{fig:2h}.
 The cross-sections are binned in 1 MeV intervals, while
the analyzing powers are binned in 5 MeV intervals.
 The tail components of the quasielastic distributions at the
lower energy transfer side are mainly due to the 
${}^{1}{\rm S}_0$ final state interaction (FSI) 
of the residual two-proton system.
 The peak positions $\omega_{\rm QES}$ 
of the quasielastic distributions 
coincide with the energy transfers 
of the corresponding 
free {\it NN} scattering indicated by the dotted vertical lines,
which is consistent with the results 
at other incident energies 
\cite{prc_69_044602_2004,prc_47_2159_1993,prc_65_034611_2002}.

 The data have been compared with the theoretical predictions obtained through
PWIA using the computer code {\sc dpn} \cite{ptp_91_69_1994}.
 These calculations have reproduced the previous 
${}^{2}{\rm H}(p,n)$ data at $T_p$ = 345 MeV and 
$\theta_{\rm lab}$ = $16^{\circ}$, $22^{\circ}$, and $27^{\circ}$, 
not only for the cross-sections but also for the polarization 
observables \cite{prc_69_044602_2004}.
 The previous data were measured at the neutron 
time-of-flight facility \cite{nima_369_120_1996}
at RCNP under a significantly low-background condition.
 Thus a comparison between the present data and 
these theoretical calculations allows us to investigate the 
reliability of the present background subtraction.
 In the calculations, the wave functions of the initial 
deuteron and the final $pp$-scattering state are generated 
by the Reid soft core potential.
 Both $S$ and $D$ states are included in the deuteron, 
and the FSI process is also included in the $pp$ scattering.
 The {\it NN} {\it t}-matrix parameterized by Bugg and Wilkin 
\cite{plb_152_37_1985,npa_467_575_1987}
is used in the impulse approximation.

 The solid curves in Fig.~\ref{fig:2h} represent the 
corresponding calculations smeared by a resolution 
function with $\Delta E$ = 7.7 MeV.
 The calculations reproduce both the cross-sections and 
analyzing powers 
around the quasielastic peak reasonably well,
but underpredict the data beyond the quasielastic peak.
 Because these calculations have reasonably reproduced 
the previous ${}^{2}{\rm H}(p,n)$ data within about 10\% 
in the present energy transfer region \cite{prc_69_044602_2004}, 
the systematic uncertainty of the present data 
around the quasielastic 
peak ($\omega_{\rm lab}$ $\lesssim$ $\omega_{\rm QES}$+20 MeV) 
is estimated to be about 10\%, whereas that 
beyond the quasielastic peak would be much larger.
 Thus, in the following discussions, we focus on the 
comparison between experimental and theoretical results 
around the quasielastic peak.
 
\subsection{${}^{3}{\rm He}(p,n)$ data}

\begin{figure}
\begin{center}
\includegraphics[width=0.7\linewidth,clip]{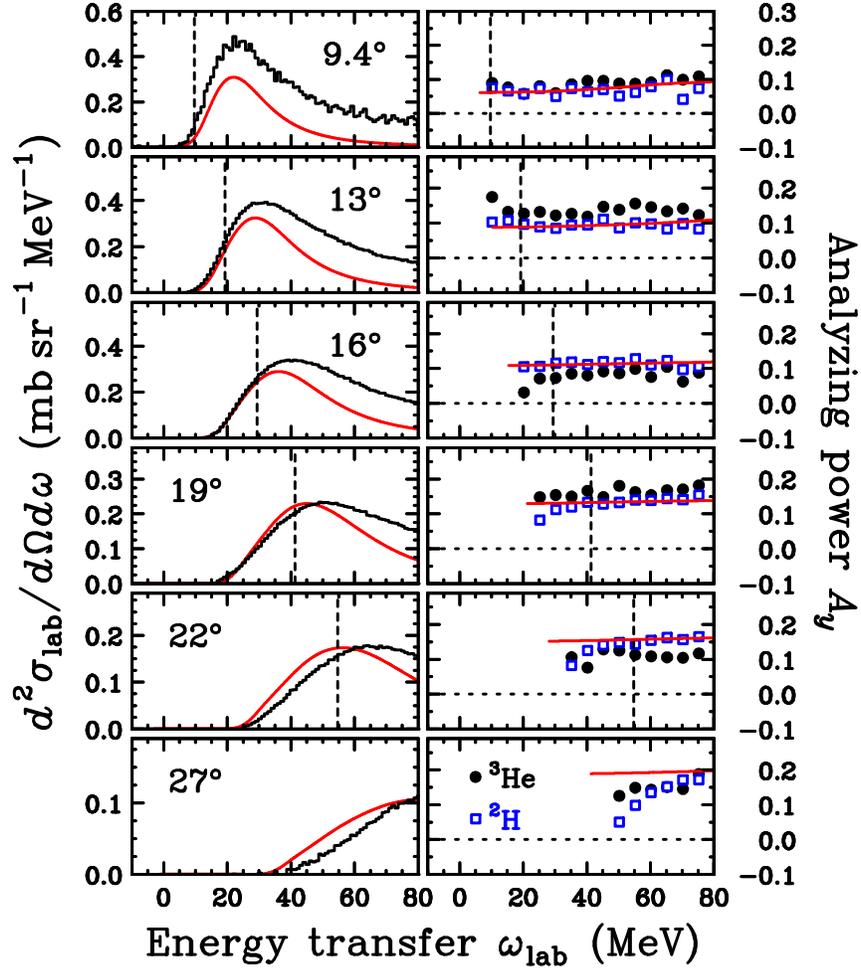}
\end{center}
\caption{(Color online)
 The cross-sections (left panels) and analyzing powers 
(right panels) 
for the ${}^{3}{\rm He}(p,n)$ reaction
at $T_p$ = 346 MeV and $\theta_{\rm lab}$ = $9.4^{\circ}$--$27^{\circ}$.
 The analyzing power data for ${}^{2}{\rm H}(p,n)$ are also shown 
by the open squares.
 The vertical dashed lines represent the energy transfers 
for the free {\it NN} scattering.
 The solid curves represent the PWIA calculations.}
\label{fig:3he}
\end{figure}

 Figure~\ref{fig:3he} shows the cross-sections 
and analyzing powers 
for the ${}^{3}{\rm He}(p,n)$ reaction 
at $T_p$ = 346 MeV and $\theta_{\rm lab}$ = 
$9.4^{\circ}$--$27^{\circ}$.
 The analyzing power data for ${}^{2}{\rm H}(p,n)$ are 
also shown by the open squares.
 The vertical dashed lines represent the energy transfers 
$\omega_{\it NN}$ for the free {\it NN} scattering.
 The analyzing powers for ${}^{3}{\rm He}(p,n)$ are in 
reasonable agreement with those for ${}^{2}{\rm H}(p,n)$. 
 Thus, the quasielastic process is expected to be dominant 
for ${}^{3}{\rm He}(p,n)$ as well.
 However, the peak position of the cross-sections is 
significantly higher than the energy transfer for the 
corresponding free {\it NN} scattering.
 The $Q$-value and Pauli principle effects are expected 
to be more 
significant than those for ${}^{2}{\rm H}(p,n)$.
 Therefore, in the following, we perform PWIA 
calculations in order to investigate these effects 
quantitatively.

\subsection{PWIA calculations for ${}^{3}{\rm He}(p,n)$}

 We performed the PWIA calculations by using the 
computer code {\sc crdw} \cite{prc_63_044609_2001}.
 The formalism for the response function is that of 
Nishida and Ichimura \cite{prc_51_269_1995}, 
and the free response function 
is employed in the present calculations.
 The single-particle wave functions were generated 
by a Woods-Saxon (WS) potential.
 The radial and diffuseness parameters were determined 
to be 
$r_0$ = 0.92 fm and 
$a_0$ = 0.38 fm, respectively, 
to reproduce the density 
distribution of ${}^{3}{\rm He}$ \cite{adndt_14_479_1974}.
 The depth of the WS potential was adjusted to 
reproduce the separation energy of the $0s_{1/2}$ orbit.
% The unbound single particle states were 
%properly treated, which is essential especially 
%to describe the resonances near threshold.
 The optimal factorization prescription 
\cite{prc_30_1861_1984,prc_33_422_1986,npa_466_623_1987,%
prc_45_1822_1992} is employed 
to model the Fermi motion of the target nucleons.
 The {\it NN} {\it t}-matrix parameterized 
by Bugg and Wilkin \cite{plb_152_37_1985,npa_467_575_1987} was used.
% In DWIA calculations, the distorted waves for incident 
%and outgoing particles 
%are calculated by using global optical potentials based 
%on Dirac phenomenology \cite{prc_73_024608_2006}.

 The solid curves in Fig.~\ref{fig:3he} represent the 
corresponding calculations.
 We focus on the energy transfer region of 
$\omega_{\rm lab}$ $\lesssim$ $\omega_{\rm QES}$+20 MeV
on the basis of the discussion in Sec.~\ref{sec:res:2h}.
 The calculations reproduce the shapes of the cross-sections reasonably well, and they also reproduce the 
analyzing powers fairly well around the peak.
 The agreement between the experimental and 
theoretical results for the peak positions 
validates the treatment of the $Q$-value and Pauli principle effects 
in the calculations.
 However, the cross-sections at small angles are significantly 
underestimated around the peak.
 This discrepancy between the experimental and theoretical 
results might be due to the 
three-nucleon resonance with isospin $T$ = $3/2$ 
\cite{prl_23_1181_1969,prc_77_054611_2008}.
 At large $\theta_{\rm lab}$ ($\ge$ $16^{\circ}$), where 
the resonance contribution is expected to be small, 
the calculations yield good descriptions of the cross-sections 
around the peak.
 Thus, we have confirmed that the quasielastic process is 
dominant around the peak region and it is described in the 
present theoretical framework without the nuclear correlations.
 Therefore, in the following, we use the same 
framework for ${}^{4}{\rm He}(p,n)$ in order to investigate 
the nuclear correlation effects in the present data.

\subsection{${}^{4}{\rm He}(p,n)$ data}

 The cross-sections and analyzing powers 
for the ${}^{4}{\rm He}(p,n)$ reaction 
at $T_p$ = 346 MeV and $\theta_{\rm lab}$ = 
$9.4^{\circ}$--$27^{\circ}$ 
are displayed in Fig.~\ref{fig:4he} as a function of 
energy transfer.
 The data for the cross-sections and analyzing powers 
are binned in 1 and 5 MeV intervals, respectively.
 In contrast to the ${}^{3}{\rm He}(p,n)$ spectra, 
there is a steep rise near the threshold at all reaction
angles.
 For small $\theta_{\rm lab}$ ($\le$ $16^{\circ}$), 
the transitions to the ground state with $J^{\pi}$ = $2^-$ 
and the first excited state with $J^{\pi}$ = 
$1^-$ of ${}^{4}{\rm Li}$ 
form a bump near the threshold.
 This bump is prominent compared with the data at 
$T_p$ = 100 MeV \cite{prc_69_024616_2004} because 
these spin-flip transitions are predominantly
excited at projectile energies around 300 MeV.
 The vertical dashed lines represent the energy transfers 
$\omega_{\it NN}$ for the free {\it NN} scattering.
 For large $\theta_{\rm lab}$ ($\ge$ $22^{\circ}$),
where the resonance contributions are expected to be small, 
the peak position of the cross-sections is significantly 
higher than the corresponding $\omega_{\it NN}$ because of 
both the large $Q$-value and nuclear correlation effects.
 In the following, we compare our data with the theoretical 
calculations by employing these effects consistently.

\subsection{PWIA calculations for ${}^{4}{\rm He}(p,n)$}

\begin{figure}
\begin{center}
\includegraphics[width=0.7\linewidth,clip]{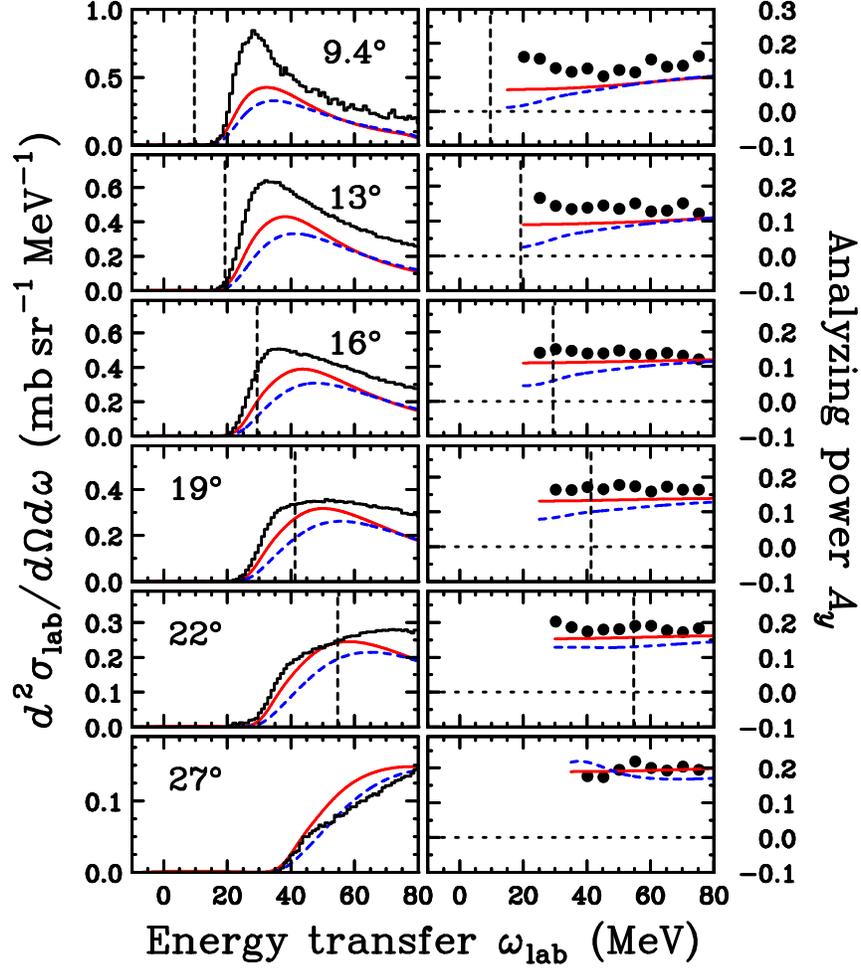}
\end{center}
\caption{(Color online)
 The cross-sections (left panels) and analyzing powers 
(right panels) 
for the ${}^{4}{\rm He}(p,n)$ reaction
at $T_p$ = 346 MeV and $\theta_{\rm lab}$ = $9.4^{\circ}$--$27^{\circ}$.
 The vertical dashed lines represent the energy transfers 
for the free {\it NN} scattering.
 The solid and dashed curves represent the PWIA calculations performed
using free and RPA response functions, respectively.}
\label{fig:4he}
\end{figure}

 The formalism of the calculations is the same as that used 
for ${}^{3}{\rm He}(p,n)$.
 The nuclear mean field is described by the WS potential, 
and its radial and diffuseness parameters were determined 
to be $r_0$ = 0.83 fm and $a_0$ = 0.33 fm, respectively, 
to reproduce the density 
distribution of ${}^{4}{\rm He}$ \cite{adndt_14_479_1974}.
 The results are shown as solid curves in 
Fig.~\ref{fig:4he}.
 The calculations significantly underestimate 
the cross-sections around the peak at small angles.
 This discrepancy is mainly due to the $L$ = 1 resonance 
contributions near the threshold, which are not described 
by the simple particle-hole excitation employed 
in the present calculations.
 For the analyzing powers, the calculations systematically yield 
smaller values than the experimental data, especially at 
small angles.
 This underestimation might also be due to the
resonance contributions in the experimental data.

 It should be noted that the observed peak positions of the 
cross-sections are significantly higher than the 
theoretical predictions at large values of $\theta_{\rm lab}$ 
($\theta_{\rm lab}$ = $22^{\circ}$ and $27^{\circ}$).
 Because the resonance contributions shift the cross-sections to lower energy transfer values, the discrepancy 
between the experimental and theoretical results 
might be indicative of nuclear correlation effects,
which are expected to be 
large for ${}^{4}{\rm He}$ \cite{prc_49_789_1994}.
 Thus we performed the PWIA calculations with the 
random phase approximation (RPA)
response functions, employing
the $\pi+\rho+g'$ model interaction 
\cite{prc_51_269_1995}.
 For the pion and rho-meson exchange interactions, 
we have used the coupling constants and meson parameters 
from the Bonn potential, which  
treats $\Delta$ explicitly \cite{pr_149_1_1987}.
 The Landau-Migdal parameters $g'$ were estimated to be 
$g'_{NN}$ = 0.65 and $g'_{N\Delta}$ = 0.35
\cite{prc_72_067303_2005,ppnp_56_446_2006} by using 
the peak position of the Gamow-Teller (GT)
giant resonance and the GT quenching factor at $q$ = 0 
\cite{prc_55_2909_1997,plb_615_193_2005}, as well as 
the isovector spin-longitudinal polarized cross-section in 
the QES process at $q$ $\simeq$ 1.7 ${\rm fm^{-1}}$ 
\cite{prc_69_054609_2004}.
 Here, we fixed $g'_{\Delta\Delta}$ = 0.5 \cite{prc_23_1154_1981}
since the $g'_{\Delta\Delta}$ dependence of the results is very weak.
 The dashed curves in Fig.~\ref{fig:4he} show the results obtained by employing 
the RPA response function.
 The nuclear correlation effects are expected to be significant 
in both the cross-sections and analyzing powers.
 We have also performed the PWIA calculations with 
the RPA response function for ${}^{3}{\rm He}(p,n)$, however, 
the nuclear correlation effects were very small.
 The cross-sections are shifted to higher energy transfer values by 
considering the nuclear correlation effects, which is due to
the hardening effects in the spin-transverse mode.
 The discrepancy of the peak positions 
at $\theta_{\rm lab}$ = $22^{\circ}$ and $27^{\circ}$ 
is resolved in part by considering the nuclear correlation
effects.
 However, it is difficult to conclude whether the nuclear 
correlations are observed in the present data because the 
resonance contributions, which are important for quantitative 
description especially at small angles, are not included 
in the present calculations.
 Thus detailed theoretical investigations including both 
the resonance and nuclear correlation effects are
required.

\section{SUMMARY AND CONCLUSION}
\label{summary}

 We have measured the cross-sections and analyzing powers 
for $(p,n)$ reactions on ${}^{3}{\rm He}$ 
and ${}^{4}{\rm He}$ 
at $T_p$ = 346 MeV and $\theta_{\rm lab}$ = 
$9.4^{\circ}$--$27^{\circ}$.
 Both data are compared with the PWIA calculations 
for quasielastic scattering.
 The calculations can reproduce the 
${}^{3}{\rm He}(p,n)$ data at 
large $\theta_{\rm lab}$ values ($\ge$ $16^{\circ}$) reasonably well.
 At small $\theta_{\rm lab}$ values ($\le$ $13^{\circ}$),
the observed cross-sections are slightly larger than the 
calculations, which might suggest the contribution 
from the $T$ = $3/2$ three-nucleon resonance.
 In contrast to the ${}^{3}{\rm He}(p,n)$ reaction, 
$L$ = 1 resonance contributions are clearly observed 
near the threshold for the ${}^{4}{\rm He}(p,n)$ reaction.
 At large $\theta_{\rm lab}$ values ($\ge$ $22^{\circ}$),
where the resonance contributions are small, 
the PWIA calculations yield reasonable descriptions 
for the cross-sections, whereas the peak positions 
are significantly lower than the experimental values.
 The observed peak shift can be explained in part by 
considering the nuclear correlations.
 However, the present data are not conclusive evidence
for the nuclear correlation effects, and call for
theoretical calculations that incorporate the proper 
description for the $L$ = 1 resonances in order to 
settle the interpretation of the present data.

\begin{acknowledgments}
 We are grateful to Professor M.~Ichimura for his 
helpful correspondence.
 We also acknowledge the dedicated efforts of
the RCNP cyclotron crew for providing a high
quality polarized proton beam.
 The experiment was performed at RCNP under 
Program Number E300.
 This research was supported in part by 
the Ministry of Education, Culture, Sports, 
Science, and Technology of Japan.
\end{acknowledgments}

\clearpage

%\subsection{}
%\subsubsection{}

% If in two-column mode, this environment will change to single-column
% format so that long equations can be displayed. Use
%\begin{widetext}
%\end{widetext}

% figures should be put into the text as floats.
% Use the graphics or graphicx packages (distributed with LaTeX2e)
% and the \includegraphics macro defined in those packages.
% See the LaTeX Graphics Companion by Michel Goosens, Sebastian Rahtz,
% entire page. There is no need to do explicit centering.

% \begin{figure}
% \end{figure}

\clearpage

\clearpage

\clearpage

\clearpage

\clearpage

\clearpage

% Surround figure environment with turnpage environment for landscape
% figure
% \begin{figure}
% \end{figure}

% tables should appear as floats within the text
% The ruledtabular enviroment adds doubled rules to table and sets a
% reasonable default table settings.
% Use the table* environment to get a full-width table in two-column
% Add \usepackage{longtable} and the longtable (or longtable*}
% environment for nicely formatted long tables. Or use the the [H]
% placement option to break a long table (with less control than 

% Surround table environment with turnpage environment for landscape
% table
% \begin{table}
% \begin{ruledtabular}
% \end{tabular}
% \end{ruledtabular}
% \end{table}

%\appendix
%\section{}

\clearpage
\bibliography{e300_4he}

\end{document}